\documentclass[reprint,aps,prl,twocolumn,10pt,superscriptaddress,nobibnotes]{revtex4-1}  
\usepackage{graphicx}
\usepackage{subfigure}
\usepackage{dcolumn}  					
\usepackage{bm}        				
\usepackage{amssymb,amsmath}
\usepackage{url}
\usepackage[
linkcolor={[rgb]{0.6,0,0}},
citecolor={[rgb]{0,0.55,0}},
urlcolor={[rgb]{0,0,0.54}},
unicode,
breaklinks,
colorlinks,
]{hyperref}
\usepackage[all]{hypcap}

\begin{document}

\title{1.5 $\mu$m lasers with sub-10 mHz linewidth}


\author{D.G.\ Matei}
\email[e-mail: ]{dan.matei@ptb.de}
\affiliation{Physikalisch-Technische Bundesanstalt, Bundesallee 100, 38116 Braunschweig, Germany}

\author{T.\ Legero}
\affiliation{Physikalisch-Technische Bundesanstalt, Bundesallee 100, 38116 Braunschweig, Germany}

\author{S.\ H{\"a}fner}	
\affiliation{Physikalisch-Technische Bundesanstalt, Bundesallee 100, 38116 Braunschweig, Germany}

\author{C.\ Grebing}
\altaffiliation{currently with TRUMPF Scientific Lasers GmbH + Co.\ KG, Feringastr.\ 10a, 85774 Unterf\"ohring, Germany}

\author{R.\ Weyrich}
\affiliation{Physikalisch-Technische Bundesanstalt, Bundesallee 100, 38116 Braunschweig, Germany}

\author{W.\ Zhang}
\affiliation{JILA, National Institute of Standards and Technology and University of Colorado, Department of Physics, 440 UCB, Boulder, Colorado 80309, USA}

\author{L.\ Sonderhouse}
\affiliation{JILA, National Institute of Standards and Technology and University of Colorado, Department of Physics, 440 UCB, Boulder, Colorado 80309, USA}

\author{J.M.\ Robinson}
\affiliation{JILA, National Institute of Standards and Technology and University of Colorado, Department of Physics, 440 UCB, Boulder, Colorado 80309, USA}

\author{J.\ Ye}
\affiliation{JILA, National Institute of Standards and Technology and University of Colorado, Department of Physics, 440 UCB, Boulder, Colorado 80309, USA}

\author{F.\ Riehle}
\affiliation{Physikalisch-Technische Bundesanstalt, Bundesallee 100, 38116 Braunschweig, Germany}

\author{U.\ Sterr}
\affiliation{Physikalisch-Technische Bundesanstalt, Bundesallee 100, 38116 Braunschweig, Germany}

\begin{abstract}
We report on two ultrastable lasers each stabilized to independent silicon 
Fabry-P\'erot cavities operated at 124~K. 
The fractional frequency instability of each laser is completely determined by 
the fundamental thermal Brownian noise of the mirror coatings with a flicker 
noise floor of $4 \times 10^{-17}$ for integration times between 0.8~s and a 
few tens of seconds. 
We rigorously treat the notorious divergences encountered with the associated 
flicker frequency noise and derive methods to relate this noise to observable 
and practically relevant linewidths and coherence times.
The individual laser linewidth obtained from the phase noise spectrum or the 
direct beat note between the two lasers can be as small as 5~mHz at 194~THz. 
From the measured phase evolution between the two laser fields we derive usable 
phase coherence times for different applications of 11~s to 55~s.

\end{abstract}
\maketitle


It is well known that frequency is the physical quantity that can be measured 
with by far the highest accuracy. 
``Never measure anything but frequency!''\ was the advice of Arthur Schawlow 
\cite{hae06}. 
The high accuracy results from the fact that the phase of a purely periodic 
signal can be measured in the simplest case by counting the zero crossings 
of the signal within a given time or with even increased accuracy by a 
phase measurement that interpolates the signal between the zero crossings. 
Hence, the generation of truly phase coherent signals over long times is the 
key to precision measurements and enabling technologies. 
In the most advanced optical atomic clocks \cite{lud15, nic15, ush15, hun16} 
pre-stabilized lasers serve as oscillators to interrogate ultranarrow optical 
transitions with linewidths of a few mHz. 
Oscillators with coherence times of tens to hundreds of seconds will allow for 
investigations of extremely small energy shifts in the clock transition, caused 
by sources such as interactions amongst atoms \cite{rey14, mar13}. 
Ultrastable oscillators beyond the state of the art will find useful 
applications in sub-mm very long baseline interferometry (VLBI) \cite{doe11}, 
atom interferometry and future atom-based gravitational wave detection 
\cite{hog16, kol16, can16}, novel radar applications \cite{ghe14}, 
the search for dark matter \cite{der14}, and deep space navigation 
\cite{gro10a}. 
Consequently, large effort has been put into the development of extremely 
coherent sources based on highly stable optical Fabry-P{\'e}rot resonators 
\cite{kes12a, hae15a, nor16a, bis13}. 
Alternative schemes are currently being investigated using cavity-QED systems 
\cite{chr15,nor16a} and spectral-hole burning in cryogenically cooled crystals 
\cite{coo15}.  


Here we report on the coherence properties of two cavity-stabilized laser 
systems operating at a wavelength of 1542~nm. 
Our systems are based on well-isolated single-crystal silicon Fabry-P{\'e}rot 
resonators, temperature stabilized at 124~K. 
For a system that has well designed locking electronics, the fractional 
frequency stability of the laser is given by the fractional stability of the 
optical length of the cavity. 
Fundamentally, the cavities' length stability is limited by statistical 
Brownian noise of the mirror coatings, substrates, and spacer \cite{num04}. 
Due to the inherently low thermal noise of crystalline silicon, the cavities' 
length fluctuations are dominated by the dielectric mirror coatings, despite 
their thickness of only a few tens of micrometers.
The cryogenic cooling of the cavities further reduces the thermal noise and 
allows for a fractional length instability of the cavities of 
$\Delta L / L \approx 10^{-17}$. 

Previously, with such a system (named Si1) we demonstrated a frequency 
instability of $1 \times 10^{-16}$ \cite{kes12a}. 
We have now set up two systems (named Si2 and Si3) where we have reduced all 
additional noise sources \cite{mat16} to a level well below the thermal noise 
limit.

In the following we describe briefly the set-up \cite{setup} and the analysis 
of the frequency stability and the phase noise. 
We subsequently derive methods to relate the dominant flicker frequency noise 
to observable and practically relevant linewidths and coherence times.


Each cavity consists of a plano-concave mirror pair employing high-reflectivity 
$\mathrm{Ta_2O_5/SiO_2}$ dielectric multilayers. 
The finesse of the TEM$_{00}$ mode of each cavity is close to 500\,000. 
The 212 mm long spacer and the mirror substrates are machined from 
single-crystal silicon \cite{kes12a}. 
The crystal orientation of the optically contacted substrates is aligned to 
that of the spacer. 
Both have the silicon $\left<111\right>$ axis oriented along the cavity axis.

The cavities are aligned vertically and are supported at three points near the 
midplane in order to minimize the impact of seismic and acoustic vibrations on 
their length stability. 
The anisotropic elasticity of silicon was used to minimize the vertical 
vibration sensitivity below $10^{-12}/$(m\,s$^{-2})$ by adjusting the azimuthal 
angle between the cavity and its tripod support \cite{mat16}. 

The cavities are placed in separate vacuum systems at a residual pressure below 
$10^{-9}$~mbar. 
The cavity temperature is stabilized to 124~K where a zero crossing of the 
coefficient of thermal expansion of silicon occurs \cite{kes12a,mat16}. 
Each system is mounted on separate optical tables, about 3~m apart.
The systems have their own active vibration isolation platforms and are 
surrounded by individual acoustic and temperature insulation boxes.
They strongly suppress individual and thus also common noise contributions to 
below the thermal noise level on timescales up to several minutes \cite{mat16}.

Commercial Er-doped distributed feedback (DFB) fiber lasers at 1542~nm 
($\nu_0 = 194.4$~THz) are frequency stabilized to the cavities using 
the Pound-Drever-Hall (PDH) method \cite{dre83}. 
Fiber-coupled acousto-optic modulators (AOM) are used for the fast servo 
allowing  locking bandwidths of around 150~kHz. 
Active residual amplitude modulation (RAM) cancellation \cite{zha14} is 
employed to keep the corresponding fractional frequency fluctuations below the 
thermal noise level of the system \cite{mat16}. 


To obtain the individual frequency instabilities of the Si2 and Si3 lasers, we 
compared them to a third ultrastable laser based on a 48~cm long ultra low 
expansion glass (ULE) cavity at 698~nm \cite{hae15a}. 
The frequency gap between the $1.5\:\mu$m Si2 system and the 698~nm ULE-cavity 
laser was bridged using a fiber-based optical frequency comb as a transfer 
oscillator \cite{tel02b,ste02a}. 
The comb introduces negligible noise that is below the thermal noise floor of 
the ULE cavity. 
Additional noise arising from the optical fibers connecting the lasers and the 
frequency comb is suppressed with active noise cancellation \cite{ma94}. 

We measured the beat frequencies `Si2 -- Si3' and `Si2 -- ULE' using 
synchronized counters \cite{kra04a}. 
The third beat frequency `Si3 -- ULE' is calculated as their difference which 
is justified since our beat measurement system does not introduce appreciable 
additional noise. 

We do not expect correlations between the ULE-cavity system, the optical 
frequency comb and the Si-systems, since they reside in three different 
rooms.   
Thus, the three difference frequencies allowed us to derive the three 
individual instabilities from a simple three-cornered hat analysis \cite{gra74} 
(Fig.\ \ref{fig:AllanVariance}). 
The relative linear frequency drift between Si2 and Si3 of about 100~$\mu$Hz/s 
(comparable with the figure reported in Ref.\ \cite{hag14}) and between Si2 and 
the ULE-cavity laser of 15~mHz/s is removed. 
 
{\begin{figure}[tb]
\centering
\includegraphics[width=0.95\linewidth]{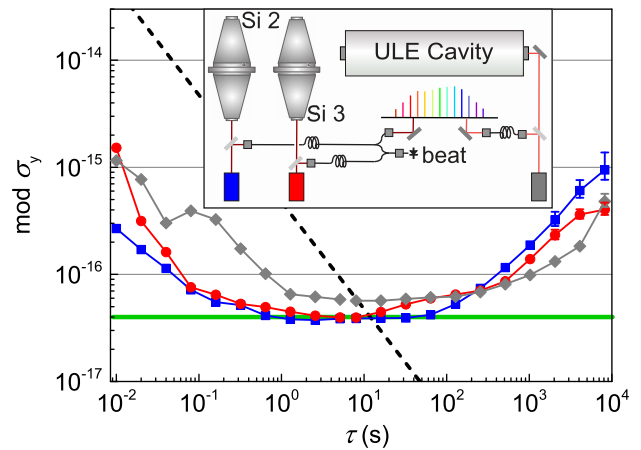}
\caption{
Modified Allan deviation for Si2 (squares), Si3 (circles) and ULE-cavity laser 
(diamonds) derived from three-cornered hat estimations. 
We used a 3.4~h dataset for $10\;\mathrm{ms}\le\tau\le4\:\mathrm{s}$ and a 
24.2~h dataset for $8\;\mathrm{s}\le\tau\le8192\;\mathrm{s}$, recorded in the 
same day.
The green line represents the expected thermal noise of the silicon cavities.
The dashed line illustrates the instability where the rms phase 
fluctuations are 1~rad for a given $\tau$ \cite{coherence}.
The intersections with the instability curves of the Si lasers result in 
coherence times of around 11~s.
Linear frequency drifts in each dataset were subtracted.
The inset shows a schematic of the measurement setup.}
\label{fig:AllanVariance}
\end{figure}}

The three-cornered hat results (Fig.\ \ref{fig:AllanVariance}) \cite{MADEV} 
indicate that for averaging times from 0.8~s up to 10~s  the instability of 
each Si-based laser system is at the expected thermal noise flicker floor of 
mod~$\sigma_y=4 \times 10^{-17}$.
This corresponds to a standard Allan deviation of about $5 \times 10^{-17}$ 
\cite{daw07}.
For short averaging times the increase in the instability is due to residual 
vibration and acoustic noise. 
At long averaging times we see the effect of slow temperature fluctuations 
affecting the cavity length and of parasitic etalons in the optical setup.

\begin{figure}[tb]
	\centering
\includegraphics[width=0.95\linewidth]{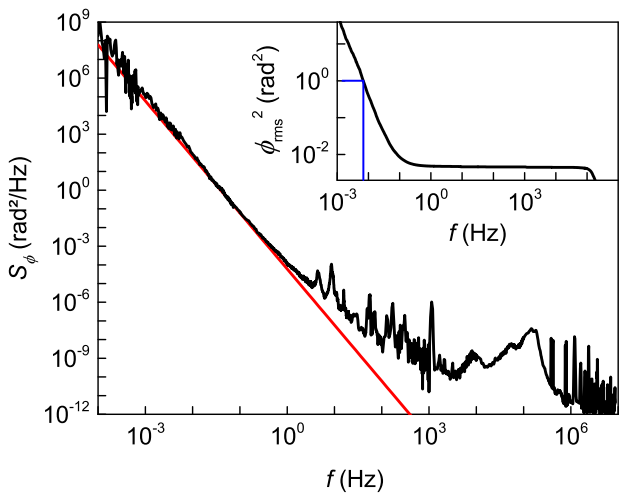}
\caption{
PSD of phase fluctuations of a Si stabilized laser, obtained as one half of the 
PSD of the Si3 -- Si2 beat. 
The red line shows the expected flicker frequency noise corresponding to the 
thermal noise at $T=124$~K. 
The inset shows the rms phase noise integrated down from 10 MHz. 
A value of 1 rad$^2$ is obtained after integrating down to 6.8~mHz (blue 
markers) leading to a FWHM linewidth of 13.6~mHz.
}
\label{fig:PSD}
\end{figure}

A more complete characterization of the noise processes is given by the power 
spectral density (PSD) of the phase fluctuations.
We have determined the phase of the beat signal from the measured in-phase and 
quadrature signal components. 
From more than 37 hours of phase data we determine the phase noise spectrum of 
a single laser down to Fourier frequencies of 0.1~mHz 
(Fig.\ \ref{fig:PSD}), modeled as 
\begin{equation}
S_\phi(f)=\nu^2_0 \sum_{k=-2}^0 h_k f^{k-2}.
\label{Eq:phasenoisemodel}
\end{equation}
From 1~mHz to 1~Hz the noise spectrum closely follows the thermal frequency 
flicker noise with $h_{-1} = 1.7 \times 10^{-33}$, in agreement with the 
expected thermal noise.
From 1~Hz to 3~kHz the seismic and acoustic perturbations above the thermal 
noise lead to a number of narrow peaks. The base line of the spectrum can be 
approximated by white frequency noise with 
$h_{0} = 3.6 \times 10^{-33} \:\mathrm{Hz^{-1}}$
consistent with the increase of the instability at short averaging times (Fig.\ 
\ref{fig:AllanVariance}).
Other possible sources such as photon shot-noise, RAM, laser power fluctuations 
are well below that level. 
At higher frequencies, the three broad peaks at 8~kHz, 60~kHz, and 150~kHz result from 
the servo loops for RAM regulation, fiber noise cancellation and PDH lock to the cavity, 
respectively.
Below 1~mHz slow temperature fluctuations lead to a random walk frequency noise 
with $h_{-2} = 4 \times 10^{-36} \:\mathrm{Hz}$, corresponding to the Allan 
deviation values above 100~s. 

In the following we use this data to derive values for laser linewidth and 
coherence time. 
Usually, linewidth and coherence time are derived from the autocorrelation 
function of the laser field with amplitude $E_0$ and center frequency $\nu_0$, 
\begin{eqnarray}\label{eq:autocorrelation}
R_E(\tau) &=& E_0^2 e^{i2\pi\nu_0\tau}
e^{-1/2 \left< (\phi(t+\tau) - \phi(t))^2 \right>}, \\
          &=& E_0^2 e^{i2\pi\nu_0\tau} \nonumber
e^{-2\int_0^\infty {S_{\phi}(f)\sin^2(\pi f \tau) df}}.
\end{eqnarray}
Flicker frequency noise and random walk frequency noise are the dominant 
noise processes in our lasers. 
In this case the laser frequency $\nu(t)$ is nonstationary and $R_E(\tau)$ is 
divergent so that no unique coherence function can be assigned. 
This also leads to divergences in the general definition of the field spectrum 
$S_E(\delta \nu)$ as the Fourier transform of the autocorrelation function 
$R_E(\tau)$ (Eq.\ (\ref{eq:autocorrelation})) and thus no uniquely defined 
linewidth exists.
Nevertheless we can derive linewidths that are closely related to the 
experimental observations. 

If a spectrum is recorded for a measurement time $T_0$ the linewidth is 
limited by the Fourier width proportional to $1/T_0$ for short measuring times 
whereas for longer measurement times the nonstationary frequency fluctuations 
broaden the line. 
In such a case a practical linewidth can be defined by the minimum. 

To elaborate this approach Bishof \textit{et al}.\ \cite{bis13} make the 
assumption that only Fourier components of the phase noise spectrum for 
frequencies $f>1/T_0$ contribute during the measurement time $T_0$.
From our phase noise model (Eq.\ (\ref{Eq:phasenoisemodel})) we obtain a 
minimal single laser linewidth of $\Delta\nu_\mathrm{FWHM}=7$~mHz for 
$T_0 = 170$~s \cite{Bishof}. 

\begin{figure}[tb]
\centering
\includegraphics[width=.95\linewidth]{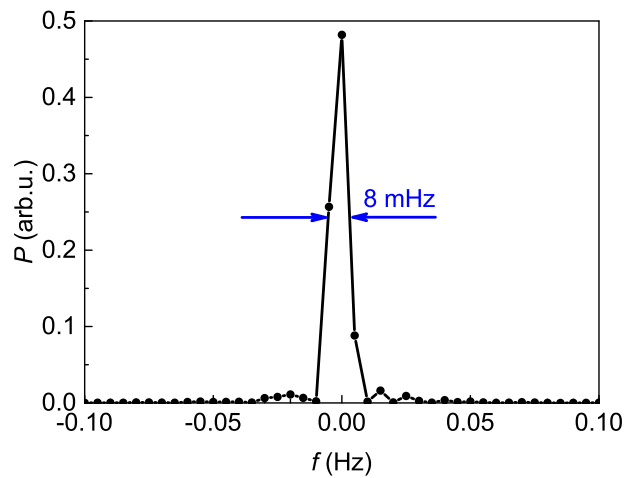}
\caption{
FFT spectrum of the beat note between lasers Si2 and Si3 (Hanning window, frequency 
resolution 7.2~mHz).}
\label{fig:FFT}	
\end{figure}

Experimentally we obtain linewidths from a fast Fourier transform (FFT) of the 
beat between the two lasers, after the beat is mixed down to a carrier 
frequency suitable for data acquisition. 
We choose 200~s measurement time to allow for sufficiently high frequency 
resolution while keeping the influence of slow frequency fluctuations small 
enough.
Experimentally, in about 43\% of the measurements \cite{FFTstat} we obtain 
full-width-half-maximum (FWHM) linewidths of the beat 
signal between 7~mHz and 14~mHz (see Fig.\ \ref{fig:FFT}), leading to 
individual linewidths $\Delta\nu_\mathrm{FWHM}$ between 5~mHz to 10~mHz, 
assuming that both lasers contribute equally to the linewidth.
This standard approach of measuring the linewidth seems to give a reasonable 
agreement with the calculated minimal linewidth of 7 mHz according to Ref. 
\cite{bis13}.

To provide a linewidth estimate that includes all fluctuations 
of the flicker frequency noise, we averaged all FFT spectra obtained from the 
data set of 37~h after first aligning their centers of mass \cite{FFTstat}.
This results in an average linewidth for a single laser of about 13~mHz for a 
measurement time of 150~s.
The difference between this longterm averaged value and the calculated minimal 
linewidth can be explained by the different ways the low-frequency cutoff is 
introduced.
If a FFT spectrum analyzer is used the spectrum is centered at the average 
frequency during the measurement time $T_0$ which corresponds to a subtraction 
of the linear phase evolution term. 
Thus significant quadratic terms still contribute to the phase excursion which 
correspond to noise at frequencies of approximately $1/2 T_0$ that is not 
included in the approximation of \cite{bis13}. 
The narrower linewidths that we have observed (Fig.\ \ref{fig:FFT}) are cases 
where the random quadratic term happened to be small.
%

Many applications are not directly sensitive to the FWHM linewidth but require 
sufficient spectral power in a narrow bandwidth $\Delta\nu_\mathrm{P}$. 
This bandwidth can be estimated by integrating the phase noise from 
high-frequencies towards zero \cite{wal75a, hal92}. 
The half bandwidth is obtained as the lower integration limit in 
\begin{equation} \label{integration}
\int_{\Delta\nu_\mathrm{P}/2}^{\infty}S_{\phi}(f)\,df
=1\,\mathrm{rad^2}\:,
\end{equation}
corresponding to the case when one third of the power is contained in the 
bandwidth $\Delta\nu_\mathrm{P}$ \cite{hal92}.
For this definition we find a value of $\Delta\nu_\mathrm{P}=14$~mHz (see inset 
of Fig.\ \ref{fig:PSD}).


For many applications it is important to provide effective coherence times of 
ultrastable oscillators.
For this purpose, depending on the particular application, different methods 
must be employed to adequately consider the nonstationary frequency. 

As an example more adequate for optical clocks we investigate a two-pulse 
Ramsey interrogation of atoms. 
There, an average frequency and frequency drift can be estimated from past 
measurements and considered in the current interrogation in order to keep the 
phase excursions $\Delta \phi$ between the two pulses sufficiently small. 

\begin{figure}[tb]
	\centering
	\includegraphics[width=0.95\linewidth]{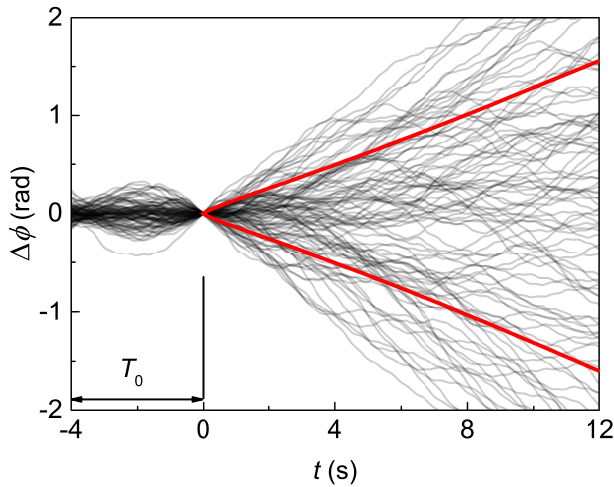}
	\caption{
	The evolution of the phase difference between the two Si lasers.
	The first 4 s segment $T_0$ is used to estimate the average frequency 
	$\overline{\nu}$ at $t = 0$~s. 
	For $t = 0 - 12$~s, the phase deviation from the expected 
	$2 \pi \overline{\nu} t$ is calculated.
	100 consecutive curves are shown with thin gray lines.
	The red lines indicate the $\pm\Delta \phi_\mathrm{rms}$ range, evaluated 
	statistically from 20\,750 curves.}
	\label{fig:PhaseTime}	
\end{figure}

We simulate such a scenario using the phase evolution of the `Si2 -- Si3' beat 
recorded for 1 day. 
We cut this dataset into short samples and fit a linear phase to the first  
4~s (i.e., observation interval $T_0$) to determine the average frequency 
$\overline{\nu}$.
The phase $2 \pi \overline{\nu} t$ is subtracted and the phase at $t=0$ is set 
to zero to obtain the phase deviation $\Delta \phi$ for $t\geq0$.
Figure~\ref{fig:PhaseTime} shows 100 of these samples, which indicate a 
time-dependent broadening.
The root-mean-square deviation $\Delta\phi_\mathrm{rms}(t)$ of the normally 
distributed phase deviation was calculated from 20\,750 samples 
($\pm\Delta\phi_\mathrm{rms}$  indicated by red lines).
The coherence is certainly lost when the phase has acquired an uncertainty of 
$\Delta\phi_\mathrm{rms} \approx \pi$~(at $t\approx 30~$s) but depending on the 
application, more restricting definitions of the coherence time are in use.  
In a more conservative way we define the coherence time as a duration in which 
$\Delta \phi_\mathrm{rms}$ has increased to 1~rad (i.e., $\sqrt{2}$~rad for the 
phase difference between the two independent lasers shown in Fig.\ 
\ref{fig:PhaseTime}).
In agreement with the value estimated from the Allan deviation 
(Fig.\ \ref{fig:AllanVariance}) \cite{coherence}, this leads to a coherence time of 11~s.
This is equivalent to saying that after 11~s in more than $99\%$ of all cases 
the actual phase excursions remain below $\pm\pi\approx 3\,\phi_\mathrm{rms}$, 
which ensures unambiguous phase tracing. 
We find that this value of 11~s represents a broad maximum in the coherence 
time when the Ramsey interrogation time varies between 4~s and 20~s 
\cite{coherence}.


Besides situations where the future phase must be predicted there are many 
applications where the average frequency can be determined in retrospect from 
the measurement itself.
Typical examples are spectral analysis, when the spectrum is centered, or the 
Rabi interrogation of atoms by single pulses, where the observed excitation 
provides the information of the average frequency during the measurement time. 
Analysis of our measured phase data shows that in this case an rms phase 
deviation of $\Delta \phi_\mathrm{rms}=1$~rad occurs at measurement intervals of about 55~s \cite{coherence}.


In conclusion, we have demonstrated the operation of two cryogenic optical 
silicon cavities at the thermal noise limit of 
mod~$\sigma_y = 4 \times 10^{-17}$. 
The light stabilized on these cavities is highly coherent, with a coherence 
time of 11~s to 55~s. 
As seen from the spectral analysis, the linewidth and implicitly the coherence 
time are mostly determined by the thermal noise level.
With this kind of laser sources we are now entering the regime where the 
frequency stability of the interrogation laser is on a par with the quantum 
projection noise limit of today's most stable optical clocks 
(e.g. \cite{sch17,cam17}).

Optimizations of the current setup would hardly bring a longer coherence time 
since we are nearing a fundamental limit. 
The only way of further improving the current performance is to decrease the 
thermal noise even further. 
One approach is to decrease the temperature, thus reducing the thermal motion 
in the system. 
For an operating temperature of 4~K the expected thermal noise would be 
$8 \times 10^{-18}$ in the modified Allan deviation. 
A comparable noise figure would be achieved by employing AlGaAs-based 
crystalline coatings, which offer a higher mechanical $Q$ factor and thus a lower 
thermal-induced noise \cite{col13, col16}. 
If both methods are implemented, the thermal noise would be reduced to the 
lower half of the $10^{-18}$ range, roughly an order of magnitude lower than 
the present level. 
To ensure that this improvement leads to an increased coherence time it is 
necessary to reduce the longterm instability for averaging times above 10~s 
(see intersection of dashed line with the thermal noise level in Fig.\ 
\ref{fig:AllanVariance}) while the present short-term instability seems to be 
sufficiently small.

Our rigorous analysis of linewidth and coherence time will be tremendously 
important when we start using this state-of-the-art laser e.g.\ for 
investigations of classical and/or quantum correlated atoms \cite{pal14}.
Achieving enhanced stability from quantum correlation (such as spin squeezing) 
will need a local oscillator that does not introduce excessive phase noise 
which can easily remove the benefit of correlation \cite{kes14}. 

\begin{acknowledgments}
This silicon cavity work is supported and developed jointly by the Centre for 
Quantum Engineering and Space-Time Research (QUEST), Physikalisch-Technische 
Bundesanstalt (PTB), the JILA Physics Frontier Center (NSF), the National 
Institute of Standards and Technology (NIST). 
This project has received funding under 15SIB03 OC18 from the EMPIR 
programme co-financed by the Participating States and from the European Union's 
Horizon 2020 research and innovation programme.
We also acknowledge support by the European Metrology Research Programme (EMRP) 
under QESOCAS. The EMRP is jointly funded by the EMRP participating countries 
within EURAMET, and the European Union.
We thank U.\ Kuetgens and D.\ Schulze for x-ray orientation of the 
spacer and mirrors, and E.\ Oelker for comments about the manuscript. 
J.\ Ye thanks the Alexander von Humboldt Foundation for support.
L.\ Sonderhouse is supported by the National Defense Science and Engineering 
Graduate (NDSEG) fellowship.
\end{acknowledgments}

\bibliographystyle{apsrev4-1}

%


\pagebreak
\begin{widetext}
\begin{center}
	\textbf{\large Supplemental Material for \\ 1.5 $\mu$m lasers with sub-10 mHz linewidth}
\end{center}
\end{widetext}
\appendix
\renewcommand{\thetable}{S\Roman{table}}
\renewcommand{\theequation}{S\arabic{equation}}
\renewcommand{\thefigure}{S\arabic{figure}}
\renewcommand{\bibnumfmt}[1]{[S#1]}
\renewcommand{\citenumfont}[1]{S#1}


\section*{Set-up: Reduction of technical noise}

The fractional frequency stability of the laser is directly related to the 
fractional stability of the optical length of the cavity. 
We therefore ensured that the external factors are reduced below the level 
given by the statistical Brownian noise. 
We address in the following the influence of temperature, laser power 
fluctuations, mechanical vibrations, and residual gas pressure fluctuations.

As temperature changes induce length fluctuations through thermal expansion, 
the operating point of the cryostat is chosen such that the cavity temperature 
precisely matches the zero-crossing point of the coefficient of thermal 
expansion (CTE) \cite{mid15}, thus reducing the impact of temperature 
fluctuations. 
These are further reduced by enclosing the cavity in two 
concentric thermal shields, with the outer one being temperature-stabilized 
using a flow of nitrogen gas and the inner one serving as a buffer. 
Care has been taken also to reduce the blackbody radiation of the environment 
reaching the cavity, by using windows that block most of it and by limiting the 
solid angle through which the radiation can enter. 
The coefficients for the heat transfer from the room temperature environment to 
the inner shield and to the cavity were measured for Si3 to be 
8(2)~$\mathrm{\mu W/K}$ and 6(2)~$\mathrm{\mu W/K}$, respectively.
For the same system, the time constants for the heat flow between cavity and 
inner shield and inner shield and active shield are 1.3 days and 6.5 days, 
respectively.
The temperature fluctuations of the cavity are thus reduced to below 1 nK for 
averaging times of a few seconds and affect the length stability 
only for times of thousands of seconds or longer \cite{mat16a}.

Fluctuations of the intracavity laser power lead to path length fluctuations 
due to heating caused by the absorbed power.
We measured a value of $1.7(2)\times 10^{-15}\mathrm{(\mu W)^{-1}}$ for both 
cavities for the proportionality coefficient between fractional frequency and 
transmitted power fluctuations.
The coefficient is small because the cavity is operated near the zero CTE point 
of the mirror substrates and due to their high thermal conductivity, and thus 
no active control of the intensity is needed. 

Vibrations transmitted to the cavity can change its dimensions, leading to 
frequency instability.
Thus we minimized the sensitivity to accelerations in all directions by 
employing a stiff holding frame. 
In addition, the sensitivity to vertical accelerations ($k_z$) was 
experimentally minimized by changing the angle between the three point support 
and the crystalline axis \cite{mat16a}. 
The acceleration sensitivities are summarized in Table \ref{table:acc_sens}.
\renewcommand{\arraystretch}{1.5}
\begin{center}
	\begin{table}[h]
		\caption{Acceleration sensitivities for the Si2 and Si3 cavities.}
		\label{table:acc_sens}
		\begin{tabular*}{\columnwidth}{@{\extracolsep{\fill}}cccc}
			\hline \hline
			& \multicolumn{3}{c}{sensitivities 
				($10^{-12} \ \mathrm{/ms^{-2}}$)}\\
			\cline{2-4}
			System & $k_x$ & $k_y$ & $k_z$\\ 
			\hline 
			Si2 & 2.5(12) & 0.7(6) & 0.4(5)  \\ 
			
			Si3 & 8.6(7) & 4.0(2) & 0.8(5) \\ 
			
			\hline \hline
		\end{tabular*} 
	\end{table}
\end{center}
Combined with the measured seismic vibrational spectrum, this ensures that the 
vibration-induced frequency noise is below the thermal-noise limit for 
averaging times above 100~ms~\cite{mat16a}.

Fluctuations in the residual gas pressure present in ion pumps \cite{rup94} 
also induce frequency instabilities by changing the refractive index of the 
residual gas between the mirrors and thus altering the optical length. 
Using ultra-high-vacuum compatible materials and keeping the ion pumps always 
in the low pressure range, we achieve a stable base pressure of $10^{-9}$~mbar. 
From the observed pressure fluctuations we estimate that corresponding 
frequency fluctuations are below $4\times10^{-17}$ for averaging times shorter 
than a few thousand seconds.

\section*{Allan Deviation}
The modified Allan deviation (mod~$\sigma_y$) is used to characterize the 
frequency stability. 
It reduces the impact of high frequency phase noise on the stability values at 
longer averaging times, as our beat signals contain phase noise at high 
frequencies that arises from the frequency comb \cite{hag13} and from laser 
noise at frequencies above the bandwidth of the PDH locks. 
The modified Allan deviation also enables to distinguish different types of 
noise, typically indistinguishable in the Allan deviation 
\cite{all81, ben15}.  

The modified Allan deviation requires frequency counters that temporally 
average the frequency fluctuations with a triangular weighting function 
(so-called $\Lambda$-counters \cite{rub05,daw07a}). 
As our counters \cite{kra04b} only approximate the \mbox{$\Lambda$-sensitivity} 
from 1~ms measurements with constant weighting function ($ \Pi$-counter), we 
additionally band-pass filter the signals with bandwidths of about 1~kHz 
to better approximate the correct sensitivity function.

\section*{Spectral Width Calculations}

The Wiener-Khintchine theorem relates the field spectrum $S_\mathrm{E}$ to the 
Fourier transform of the field autocorrelation function 
\begin{equation}
R_\mathrm{E}(\tau) = \langle E(t+\tau) E^*(t) \rangle.
\end{equation}

For a field $E(t)=E_0 e^{2\pi i \nu_0 t} e^{i\phi(t)}$ with average frequency 
$\nu_0$ and random phase $\phi(t)$ this autocorrelation function can be 
expressed as
\begin{equation}
\label{eq:ACF_phi}
R_\mathrm{E}(\tau) = E_0^2 e^{2\pi i \nu_0 \tau}
\exp \left(-\tfrac{1}{2} \Delta\phi_\mathrm{rms}^2(\tau) \right)
\end{equation}
where we have used the root-mean-square (rms) phase increment
\begin{equation}
\label{eq:Dphirms}
	\Delta\phi_\mathrm{rms}^2(\tau) = \langle 
	\left(\phi(t+\tau)-\phi(t)\right)^2 \rangle.
\end{equation}
The phase increment can be calculated with a sensitivity function
\begin{equation}
h^{(\tau)}(t) = \delta(t-\tau) - \delta(t)
\end{equation}
as
\begin{equation}\label{eq:varphi}
\phi(t+\tau)-\phi (t) = \int \phi(t+t') h(t') dt',
\end{equation}
using the Dirac delta function $\delta(t)$.
With the help of Parseval's theorem, the rms value of this convolution can be 
expressed through the power spectral density of phase fluctuations 
$S_{\phi}(f)$ as 
\begin{eqnarray}\label{eq:ACF_Sphi}
\Delta\phi_\mathrm{rms}^2(\tau) 
&=& \int_0^\infty S_{\phi}(f) |H(f)|^2 df \\
&=&  4\int_0^\infty S_{\phi}(f) \sin^2(\pi f \tau) df
\end{eqnarray}
where we use the Fourier transform of the sensitivity function $h(t)$
\begin{equation}
H(f) = \int_0^{T_0} h(t) \exp(2 \pi i f t) dt. 
\end{equation}

With the power spectral density of frequency fluctuations 
$S_{\nu}(f) = f^2 S_{\phi}(f)$ the corresponding autocorrelation function reads
\begin{equation}
\label{eq:ACF_Snu}
R_E(\tau) = E_0^2 e^{2\pi i \nu_0\tau} 
	\exp \left({-2\int_0^\infty {S_{\nu}(f)\frac{\sin^2(\pi f \tau)}{f^2} 
	df}}\right). 
\end{equation}

However, for frequency noise processes $S_{\nu} \propto f^k$ that are diverging 
towards zero frequency with $k\leq -1$, this autocorrelation function is 
diverging.  
This is due to the fact that the phase difference, expressed by the average 
frequency $\overline{\nu}_{\tau}(t)$ in the interval $[t,t+\tau]$ 
\begin{eqnarray}
\Delta\phi(t) 
&=& \phi(t+\tau)-\phi(t) \\
&=& 2\pi\tau \overline{\nu}_{\tau}(t)
\end{eqnarray}
is nonstationary, so the expectation value needed to define the 
autocorrelation function $R_\mathrm{E}$ does not exist.

A similar problem appears when trying to use the classical frequency variance 
$\langle {\overline{\nu}(t)}^2 \rangle$ 
to describe the stability of oscillators in time domain \cite{all87}. 
There the Allan variance $\sigma_{\nu}^2$ is now widely used instead to 
describe the stability of such sources,  
which circumvents the divergence of the classical variance by taking the 
variance between successive average frequencies:
\begin{equation}
\sigma_{\nu}^2(\tau) = \tfrac{1}{2} \langle 
\left(\overline{\nu}_{\tau}(t+\tau)-\overline{\nu}_{\tau}(t)\right)^2 \rangle.
\end{equation}

\subsection{Low-frequency cutoff methods}

As only finite observation times $T_0$ are used in any real experiment, it is 
common to avoid the divergence by introducing low-frequency cutoffs 
$f_\mathrm{co}$ in Eq.~(\ref{eq:ACF_Snu}). 
In the work of Stephan \emph{et al}.\ \cite{ste05} a cutoff at 
$f_\mathrm{co}=1/\tau$ is introduced. 
For pure flicker noise $S_{\nu} = h_{-1} f^{-1}$ this approach leads to a 
Gaussian line profile and an effective FWHM linewidth 
$\Delta \nu =  0.3537 \nu_0 \sqrt{h_{-1}}$
independent of observation time.

Bishof \emph{et al}.\ \cite{bis13a} introduce a cutoff at $f_\mathrm{co}=1/T_0$, 
which leads to a linewidth that depends on the observation time $T_0$, and its 
minimum is used as the effective linewidth.
To include also the Fourier width due to the limited observation time, 
windowing functions $w(t)$ are employed in the finite-length Fourier transform 
\cite{har78a}, leading to a spectrum of
\begin{equation}
S_\mathrm{E}(f) 
=  \int_0^\infty W(\tau) R_E(\tau) \cos(2\pi  f \tau) d\tau.
\end{equation}
Here the weighting function for the autocorrelation function $W(\tau)$ is the 
convolution of the initial weighting function with itself 
$W(\tau)= (w\ast w)(\tau)$. 
E.g.\ in the case of a rectangular window function of duration $T_0$ it is
\begin{equation}
W(\tau) =  (1-|\tau|/T_0).
\end{equation}

\subsection{Practical spectral measurements}
The above methods do not directly correspond to practically employed spectral 
measurements. 
One widely used method to measure an effective linewidth is the spectral 
analysis of the beat signal between two similar oscillators during a limited 
measurement duration $T_0$ using spectrum analyzer \cite{ban16}, often based on 
a Fast Fourier Transform (FFT) of the signal.
Here naturally only the width is recorded, while the average frequency of the 
beat is manually tracked to keep the signal within the observation bandwidth, 
which compensates for the nonstationary frequency of flicker noise. 
E.g.\ for a spectral measurement of duration $T_0$, the average frequency can 
be determined from the spectrum itself as the central frequency of the observed 
spectral feature. 

Mathematically, this means that no longer the complete phase evolution 
$\phi(t)$ is analyzed over infinite durations. 
Instead the expectation value of finite duration spectra from 
$\phi^\mathrm{(cor)}(t)$ are considered, where the phase 
$\phi^\mathrm{(av)}(t)$ due to the average frequency $\nu^{\mathrm{av}}$ is 
subtracted from each individual spectrum.
Thus the variance of the phase increments in Eq.~(\ref{eq:Dphirms}) is not 
taken from the real laser phase but for the phase increments corrected by an 
average phase increment $2 \pi \nu^\mathrm{(av)} (t_2-t_1)$ during the 
observation time with an average frequency $\nu^\mathrm{(av)}$.    

Steck \cite{ste16} uses a weighted averaging depending on $\tau$ and $T_0$ to 
obtain a FWHM as function of $T_0$. 
Its minimum for flicker noise $S_{\nu}(f) = h_{-1}/f$ is
$0.5 \nu_0 \sqrt{h_{-1}}$ at $T_0= 14 \nu_0 / \sqrt{h_{-1}}$.

It should be noted that, due to this subtraction, the corrected phase increment 
in the observation interval in general is no longer invariant to time 
translation, but now depends on the two times: $\Delta\phi(t_1, t_2)$. 
The finite-length spectrum (periodogram) with window function $w(t)$ is given 
as absolute squared Fourier transform of the signal:
\begin{eqnarray}
S_\mathrm{E}(f) &=& |\mathcal {F}_E(f)|^2  \label{eq:fieldspectrum} \\
&=&  \int_0^{T_0} \int_0^{T_0} w(t_1) w(t_2) 
	 e^{-\tfrac{1}{2} \langle \Delta\phi^2(t_1, t_2) \rangle} \nonumber \\
	 &\cdot& \cos(2 \pi f (t_2-t_1)) dt_1 dt_2 \nonumber.
\end{eqnarray}

In the simplest way, the average frequency can be calculated from the phase 
increment during the interval $[t,t+T_0]$:
\begin{equation}
\label{eq:f_av}
\phi^{(\mathrm{av})} = 2\pi\tau \overline{\nu}(t) = \phi(t+T_0)-\phi(t). 
\end{equation}

The sensitivity function that corresponds to this interpolation by the average 
frequency is
\begin{equation}
h^{(\mathrm{int})}(t_1,t_2) = (t_2-t_1)/T_0 \: (\delta(t-T_0) - \delta(t)),
\end{equation}
with the corresponding Fourier transforms
\begin{eqnarray}
H^{(\tau)}(t_1,t_2,f) &=& 	e^{2\pi i f t_2} - e^{2\pi i f t_1}, \\
H^\mathrm{(int)}(t_1,t_2,f)  &=& (t_2-t_1)/T_0 \:	(e^{2\pi i f T_0}-1), \\
H^{(\mathrm{diff})}(t_1,t_2,f)  &=& H^{(\tau)}(t_1,t_2,f) - 
H^\mathrm{(int)}(t_1,t_2,f).\quad 
\end{eqnarray}

\begin{figure}[tb]
	\centering
	\includegraphics[width=0.92\linewidth]{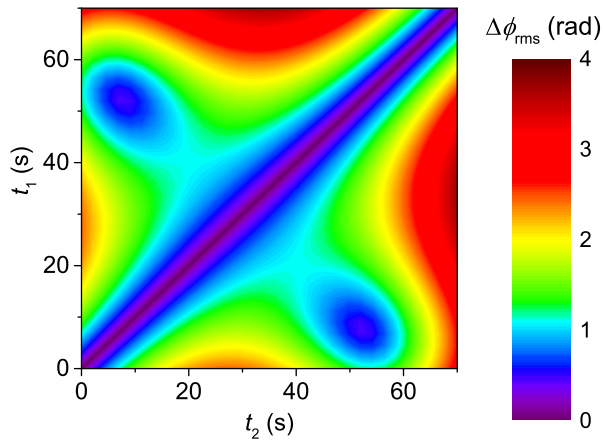}
	\caption{
		Phase deviation $\Delta\phi_\mathrm{rms}(t_1,t_2)$ for the 
		experimentally observed spectrum of frequency fluctuations, a duration 
		of the observation interval $T_0=60~$s and a fit with rectangular 
		weighting function. 
	}
	\label{fig:VarPhi}
\end{figure}

The rms phase deviation $\Delta\phi_\mathrm{rms}(t_1,t_2)$ is thus 
expressed with the help of Parseval's theorem as:
\begin{equation}
\Delta\phi_\mathrm{rms}^2(t_1,t_2) =  
	\int_0^\infty S_\phi(f) \left|H^{(\mathrm{diff})}(t_1,t_2,f) \right|^2 df.
\end{equation}
An example of $\Delta\phi_\mathrm{rms}(t_1,t_2)$ using a least-squares fit 
of the laser phase with $T_0=60$~s is shown in Fig.\ \ref{fig:VarPhi}. 
The figure indicates clearly that the phase deviation does not simply depend on 
the difference $t_1-t_2$. 
The chart provides the full information of rms phase deviations for different 
measurement scenarios. 
The lower edge of the diagram, showing $\Delta\phi_\mathrm{rms}(0,t_2)$, 
applies to Rabi interrogation and  will be discussed below in connection to the 
experimental data shown in Fig.\ \ref{fig:Phase_Rabi}. 
The data relevant for Ramsey interrogation are visible in the stripe 
($t_1=60$~s, $t_2>60$~s). 
The diagram and the underlying calculations provide also the relevant 
information if there is a a gap between the initial observation interval and 
the subsequent prediction.

With the rms phase deviation $\Delta\phi_\mathrm{rms}(t_1,t_2)$ and a window 
function $w(t)$ one calculates the field autocorrelation for $\tau>0$
\begin{equation}
R_\mathrm{E}(\tau) = E_0^2 
\int_0^{T_0-\tau} 
	w(t)\, w(t+\tau)\, e^{-\tfrac{1}{2} \Delta\phi_\mathrm{rms}^2(t,t+\tau) } dt
\end{equation}
and the spectrum according to Eq.\ (\ref{eq:fieldspectrum}).

A better approximation to the average frequency is a least-squares fit to the 
laser phase, weighted by the window function $w(t)$ of the FFT. 
Without loss of generality we consider the interval $[0,T_0]$. 
For the fit we use a sum of orthogonal polynomials $\Pi_k(t)$ over the interval 
$[0,T_0]$ with weight $w(t)$.
For constant weight $w=1$ these are the shifted Legendre polynomials.  
Then in the least-squares sense the phase is approximated by
\begin{equation}
\phi^\mathrm{fit}(t) = \sum_{k=0}^N c_k \Pi_k(t) ,
\end{equation}
with coefficients
\begin{equation}
c_k = \int_0^{T_0} w(t) \Pi_k(t) \phi(t) dt.
\end{equation}
Thus these coefficients can be expressed as convolution between the phase 
$\phi(t)$ and a kernel $w(t) \Pi_k(t)$, and the variance of the corrected 
phase can be expressed with the help of Parseval's theorem through the product 
of $S_\phi$ and the square of the absolute value of the Fourier transform 
$\mathcal F(w(t) \Pi_k(t))$.

\begin{figure}[tb]
	\centering
	\includegraphics[width=0.95\linewidth,height=57mm]{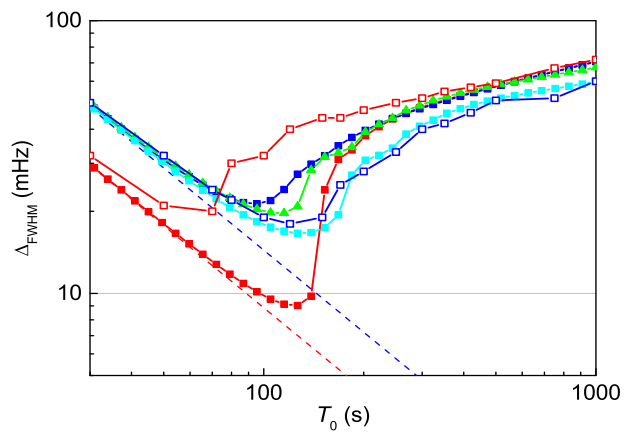}
	\caption
{FWHM beat linewidth $\Delta_\mathrm{FWHM}$ as a function of observation time 
$T_0$.
Filled symbols: Calculations using the modeled phase noise spectrum and 
different methods to deal with the low-frequency divergence:
with a cutoff at $f = 1/T_0$ \cite{bis13a} and rectangular window (red squares) 
or Hanning window (blue squares), with subtraction of phase frequency from 
weighted linear fit and Hanning window  (green triangles) and with weighted 
quadratic fit and Hanning window (cyan squares).  
		Open symbols: Linewidths obtained by averaging FFT spectra obtained
with different window functions: rectangular (red squares) and Hanning (blue 
squares).
The dashed lines indicate the respective Fourier limits of the rectangular 
(red) or Hanning window (blue).
	}
	\label{fig:BeatFWHM}
\end{figure}

\begin{figure}[b]
	\centering
	\includegraphics[width=0.95\linewidth,height=57mm]{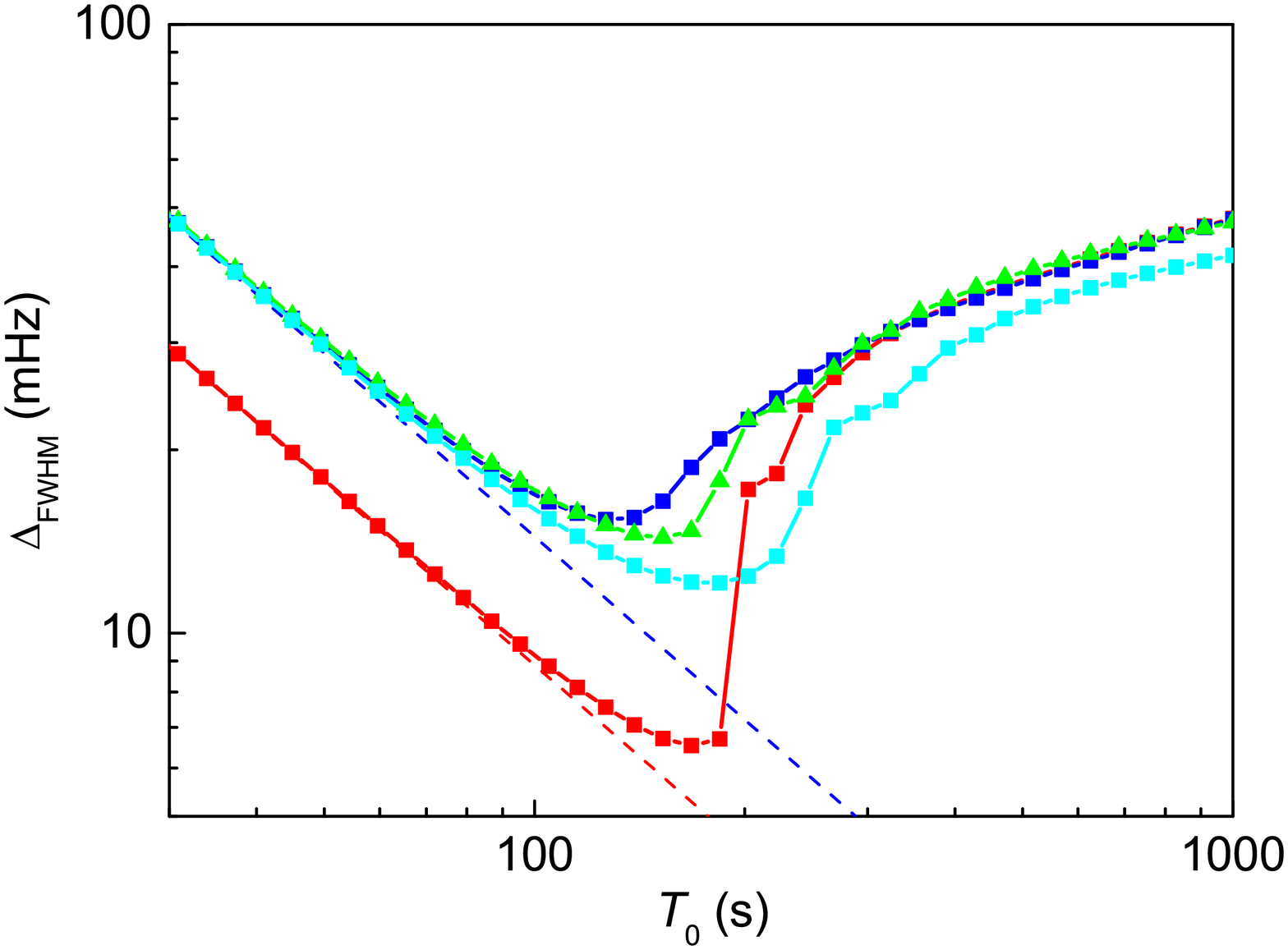}
	\caption{
	Single laser FWHM linewidth $\Delta_\mathrm{FWHM}$ as a function of 
	observation time $T_0$ calculated from the modeled phase noise spectrum. 
		For plot legend see caption of Fig.\ \ref{fig:BeatFWHM}. 
	}
	\label{fig:SingleLaserFWHM}
\end{figure}

The rectangular (constant) weighting window and the Hanning window 
\cite{har78a} 
\begin{equation}
w(t) = 1-\cos(2\pi t/T_0) 
\end{equation}
are widely used in FFT spectral analysis.
For both window functions we have calculated the phase variance (Fig.\ 
\ref{fig:VarPhi}) and the linewidth as function of the measurement interval 
length $T_0$ (Fig.\ \ref{fig:BeatFWHM} (filled symbols) and 
Fig.\ \ref{fig:SingleLaserFWHM}). 
If only the linear phase is fitted ($N=1$), we expect to obtain the linewidth 
of the averaged spectra.
If the fit also includes a quadratic term ($N=2$), we expect to find the 
minimum observed linewidth, as during these measurements the actual frequency 
drift (i.e.\ the quadratic phase) was close to zero. 

\subsection{FFT statistics}
\begin{figure}[tb]
	\centering
	\includegraphics[width=0.95\linewidth,height=58mm]{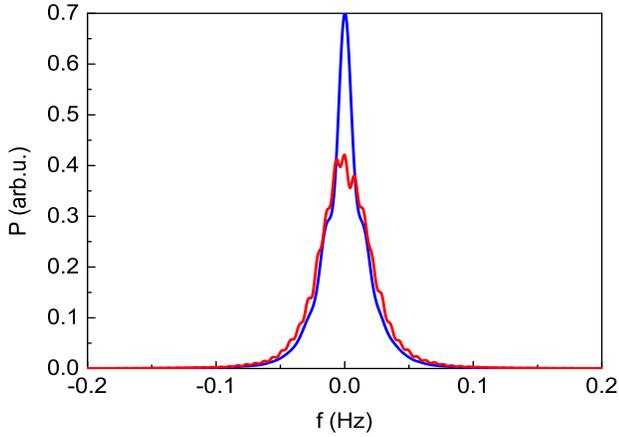}
	\caption{
		Averaged FFT spectrum of the beat of the two lasers obtained from a 
		phase measurement of 37~h by averaging all spectra obtained from 150~s 
		intervals with a rectangular window (red line) and Hanning window (blue 
		line). 
		The increased frequency resolution results from zero-padding the 
		data before the FFT calculation.
	}
	\label{fig:FFT_avg}
\end{figure}

For a complete characterization of the linewidth measurements with FFT we used 
the 37~h phase record employed for calculating the phase noise spectrum. 
The data was broken up in adjacent equal-length segments and a FFT spectrum was 
obtained for each of them.
The spectra were aligned on the frequency axis with their centers of mass 
at 0~Hz.
For a finer alignment, the resolution bandwidth of the FFT was 
increased artificially by zero-padding the segments up to eight times their 
length.
For each frequency the mean value from all spectra was calculated, resulting in 
an averaged spectrum as displayed in Fig.\ \ref{fig:FFT_avg}.
When varying the length of the segments we obtain the data shown in 
Fig.\ \ref{fig:BeatFWHM} with open symbols.
It results that the optimum interval length for obtaining a minimal linewidth 
lies between 120~s and 150~s.

Since the individual FFT spectra usually have irregular shapes to which no 
analytic peak function can be assigned, we use an empirical approach in 
estimating their linewidths.
First the maximum value was determined.
Then the maximum was approached from both ends of the spectrum until the 
half-value was encountered. 
The difference between the two frequency values was then taken as the FWHM 
value.
Using the data from Fig.\ \ref{fig:FFT_avg} we obtain a linewidth of 19~mHz 
(for a Hanning window) of the beat, which results in a single-laser average 
linewidth of about 13~mHz.
\begin{figure}[tb]
	\centering
	\includegraphics[width=0.95\linewidth,height=73mm]{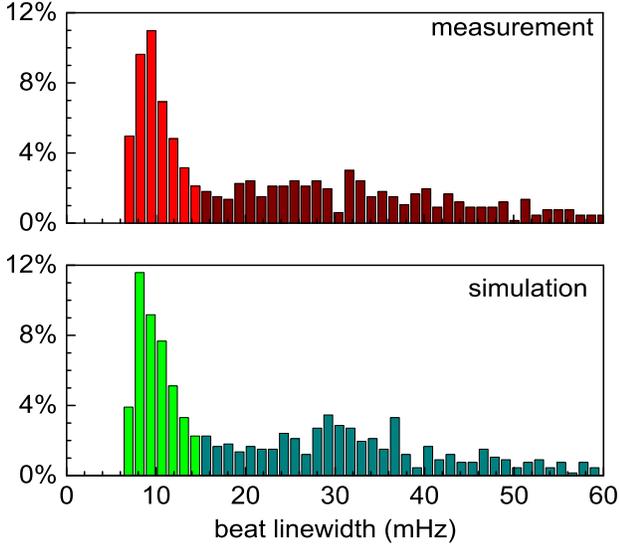}
	\caption{
		Histogram of spectral linewidths for measured beat data (upper graph) 
		and simulated flicker frequency noise (lower graph) corresponding to 
		200~s segments obtained from a 37~h record. 
	}
	\label{fig:FFT_statistics}
\end{figure}
We also calculate the distribution of linewidths over the time span of 37~h.
The result is shown in the upper graph from Fig.\ \ref{fig:FFT_statistics} for 
an interval of 200~s. 
This corresponds to the measurement with a FFT analyzer shown in the main text.
The highlighted part represents the beat linewidths below 14~mHz 
which amounts to 43\% of all measurements.
For comparison, the same analysis for a simulated pure flicker frequency noise 
is shown in the lower graph.
The similarity of the two histograms confirms once again that in this time 
range the behavior of the lasers is essentially described by a flicker 
frequency noise and that the broad distribution of linewidths is intrinsic to 
$1/f$ noise and not due to additional technical perturbations. 
\section*{Coherence Time}
The coherence time $T_\mathrm{co}$ can be defined \cite{sal91} 
as the time where the autocorrelation function $R_E(\tau)$ has fallen to a 
certain fraction (e.g.\ $1/2$ or $1/e$) of its value at $\tau=0$. 
According to Eq.\ (\ref{eq:ACF_phi}) the definition of coherence time 
$T_\mathrm{co}$ by 
$R_\mathrm{E}(T_\mathrm{co}) = 1/e$ 
corresponds to $\Delta\phi^2(T_\mathrm{co}) = 2~\mathrm{rad}^2$.

A relation between coherence time and FWHM linewidth $\Delta \nu$ can be found 
in \cite{sal91} which gives 
$T_\mathrm{co} = 1/\Delta \nu$ for rectangular, 
$T_\mathrm{co} = 0.32/\Delta \nu$ for Lorentzian 
and $T_\mathrm{coh} = 0.66/\Delta \nu$ for Gaussian spectra. 
\subsection*{Coherence time for Ramsey interogations}
\begin{figure}[tb]
	\centering
	\includegraphics[width=0.95\linewidth,height=60mm]{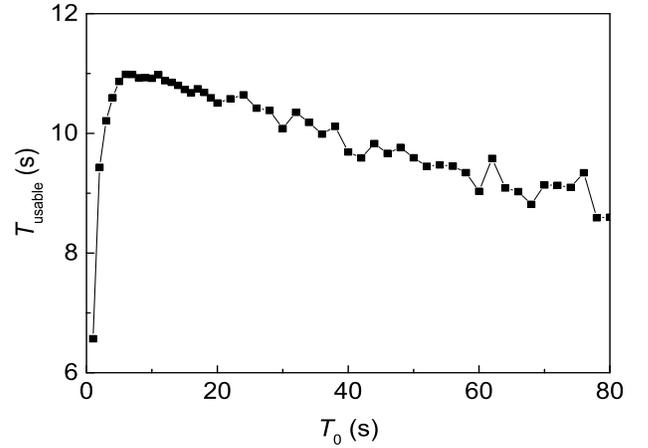}
	\caption{Usable time $T_\mathrm{usable}$ until the phase deviation 
	$\Delta \phi_\mathrm{rms}$ between predicted and actual phase for a single 
	laser exceeds 1~rad calculated for different durations $T_0$ of the 
	preceding observation interval.
	\label{fig:coherence_Ramsey}}	
\end{figure}

For Ramsey spectroscopy, the frequency of the interrogating laser needs to be 
measured in advance. 
Depending on the observation time $T_0$ chosen to measure the frequency, 
the quality of the predicted phase evolution and thus the usable time 
$T_\mathrm{usable}$ until the phase deviation between predicted and actual 
phase for a single laser exceeds $\Delta \phi_\mathrm{rms}=$ 1~rad may change. 
Using the measured phase data, we have calculated $T_\mathrm{usable}$ as a 
function of the preceding observation time (Fig.~\ref{fig:coherence_Ramsey}). 
The steep decrease of $T_\mathrm{usable}$ towards short observation times 
results from the increased influence of high frequency noise that leads to a 
poor predictability.  
The decrease at long observation times is a property of the $1/f$ frequency 
noise.
The optimal observation time $T_0 \approx 4$~s leads to a 
$T_\mathrm{usable}$ of 11~s which is the practical coherence time for the 
application in two-pulse Ramsey interrogation of atoms. 
\subsection*{Coherence time for Rabi interrogations}
For Rabi interrogation the frequency of the laser does not need to be known in 
advance but can be determined retrospectively from the result of the 
measurement. 
To cover the fluctuating average frequency during the Rabi interrogation, 
in analogy to the principle of a FFT spectrum analyzer, we consider an array of 
atomic ensembles at slightly differing center frequencies. 
Then the location of the maximum excitation probability provides information 
about the average frequency during the interrogation. 
In this scenario, a useful coherence time may be defined as the maximum time 
where the on-resonance excitation probability is not reduced too much from 
unity. 
If we linearize the optical Bloch equation, in analogy to the Strehl ratio in 
wave optics \cite{mar82} the maximum excitation probability is approximated by 
\begin{equation} \label{eq:marechal}
P_\mathrm{max} = e^{-\langle \Delta\phi^2_\mathrm{rms} \rangle}.
\end{equation}
We simulate this condition by dividing a 23~h phase evolution data interval in 
equal-length intervals of duration $T_0$ which corresponds to the Rabi 
interrogation time. 
We obtain the average frequency in each interval by a least-squares fit to the 
phase data and subtract it from the measured phase evolution. 

\begin{figure}[tb]
	\centering
	\includegraphics[width=0.95\linewidth,height=57mm]{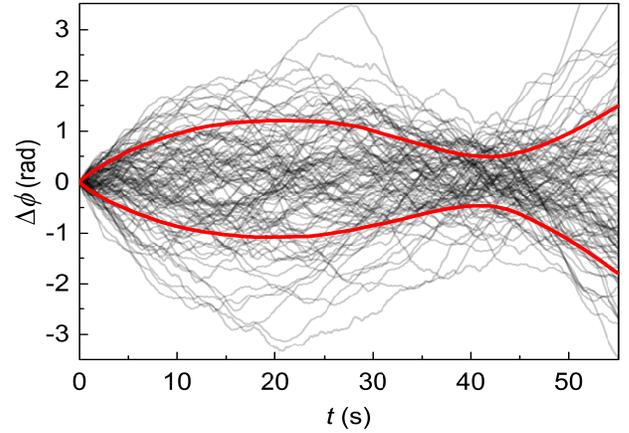}
	\caption{
	Measured phase fluctuations of the beat of 100 consecutive curves (thin 
	gray lines) with the phase from the average frequency $\overline\nu$ over 
	$[0, T_0]$ with $T_0= 55~$s of each curve subtracted.
	The red lines indicate the $\pm\Delta \phi_\mathrm{rms}(t)$ range, 
	calculated from 1\,500 curves.
	For this length of the temporal average of the rms fluctuations 
	$\langle \Delta\phi^2_\mathrm{rms} \rangle$ is 1~rad.
	}
	\label{fig:Phase_Rabi}	
\end{figure}

Fig.~\ref{fig:Phase_Rabi} shows the remaining phase $\Delta \phi$ of the beat 
signal between Si2 and Si3 for an interval length of $T_0=55$~s for 100 
consecutive intervals. 
The accumulated rms phase $\Delta\phi_\mathrm{rms}(t)$ during each interval is 
shown in red. Note that due to fitting all these curves depend on the chosen 
interval length $T_0$. 
Because the spectrum of the phase fluctuation strongly increases towards low 
frequency, residual phase fluctuations after removing the fit contains a strong 
component with frequency $1/2T_0$ which leads to the peculiar shape of 
$\Delta\phi_\mathrm{rms}(t)$. 
To apply Eq.\ (\ref{eq:marechal}) we have to take the rms value of the red 
curve within $[0,T_0]$. 
For the conditions of Fig.~\ref{fig:Phase_Rabi} we find for the beat $\langle 
\Delta\phi^2_\mathrm{rms} \rangle = 1$~rad$^2$. 
Assuming identical laser performance the rms phase of each laser is then given 
by $\Delta \phi^2_\mathrm{rms} = 1/2$~rad$^2$ and $P_\mathrm{max} = 0.6$ which 
would be still a useful excitation probability. 
Thus this interval length of $T_0= 55$~s can be considered as practically 
relevant coherence time for Rabi interrogation.

It is interesting why for Ramsey and Rabi interrogation of quantum systems 
with a highly stable laser governed by flicker noise the derived coherence time 
largely differs as 11~s and 55~s, respectively. 
One can understand this difference by the amount of prior information available 
for the interrogation. 
In Ramsey interrogation one has to estimate the mean frequency of the laser 
(and possibly its drift rate) from the previous evolution of these quantities. 
Flicker noise in the laser field inevitably means that the real phase evolution 
differs from the predicted one during the interrogation period. 
In Rabi interrogation, less prior information is needed since the measurement 
by itself gives the actual information about the mean frequency.
Here, the situation is pretty much the same as with the linewidth measurement 
with a spectrum analyzer. 
There any drift of the mean frequency will only lead to a shift of the acquired 
field spectrum. Thus both the information on the average frequency and the 
linewidth is available afterwards. 
In contrast, the Ramsey interrogation can only deliver information about this 
mean frequency as long as the phase difference between atomic coherence and 
laser field remains within $\pm \pi$, which puts much more severe constraints 
on the quality of the prediction.

\subsection*{Relation between Allan deviation and coherence time}

To predict the future phase under the condition of nonstationary frequency 
fluctuations, as in the case of  flicker and random walk frequency noise, the 
future frequency needs to be estimated using an average frequency from the past 
values. 
As a convenient way for extrapolation over a duration $T_0$ in an interval 
$\left[t,t+T_0\right]$, the average frequency from the preceding interval with 
the same duration $\left[t-T_0,t\right]$ can be used.
The variance between successive frequencies is used in the definition of the 
Allan deviation
\begin{eqnarray}
\sigma_y^2(T_0) 
	&=& \frac {1}{\nu_0^2} \langle \frac{1}{2} \left(\bar{\nu}_{i-1} - 
	\bar{\nu}_i\right)^2 \rangle \\
	&=& \frac {1}{4\pi^2 T_0^2\nu_0^2} \langle \frac{1}{2} 
	\left(\Delta\phi_{i-1} - \Delta\phi_i\right)^2 \rangle , 
\end{eqnarray}
where $\nu_0$ denotes the average frequency, $\overline{\nu_i}$ the average 
frequency over the interval of duration $T_0$, and $\langle.\rangle$ the 
expectation value.
If the frequency is adjusted to the frequency of the preceding interval, this 
corresponds to $\Delta\phi_{i-1} = 0$, and thus
\begin{equation}
 	\Delta\phi^2_\mathrm{rms} = 8\pi^2\nu_0^2 T_0^2 \sigma_y^2(T_0) = \langle 
 	\Delta\phi_i^2 \rangle .
\label{eq:varphi_adev}
\end{equation}

This equation allows to estimate the coherence time from the measured 
frequency stability as shown in \mbox{Fig.\ 1}. 
A phase deviation of $\Delta\phi_\mathrm{rms}=1$~rad corresponds to an Allan 
deviation of
\begin{equation}
 	 \sigma_y(T_0) = \frac{1}{2 \sqrt{2}\pi\nu_0} T_0^{-1}.
	\label{eq:sigma_y}
\end{equation}
The coherence time $T_\mathrm{co}$ is then defined by the intersection with the 
individual frequency stability of the lasers.

The Allan deviation plot (Fig.\ 1) shows that the coherence time for Si2 and 
Si3 is about 11~s. 
Please note that $T_\mathrm{co}$ is not limited by the short term stability of 
the lasers, but only by the flicker floor, given by the thermal noise of the 
cavities.  

On the other hand, if the average frequency $\overline{\nu}$ is determined from 
the measurement itself using 
$\overline\nu = (\phi(t+T_0)-\phi(t))/2\pi T_0$, 
the maximum phase excursion is expected at the midpoint of the interval 
$t+T_0/2$ and its variance is given by
\begin{equation}
\Delta\phi_\mathrm{rms}^2 =   
 \langle \left( \phi(t+T_0/2) - \frac{\phi(t+T_0)+\phi(t)}{2} 
 \right)^2\rangle,
\end{equation}
which is
\begin{equation}
\Delta\phi_\mathrm{rms}^2 =   
 1/4 \cdot (2\pi T_0/2)^2 \langle \left( \overline{\nu}_i - 
 \overline{\nu_{i+1}} \right)^2\rangle,
\end{equation}
where $\overline{\nu_i}$ and $\overline{\nu_{i+1}}$ denote the average 
frequency over the intervals $[t,t+T_0/2]$ and $[t+T_0/2,t+T_0]$.

This can be expressed by the Allan deviation as:
\begin{equation}
\Delta\phi_\mathrm{rms}^2 = 1/2 \cdot \pi^2 T_0^2 \nu_0^2 \sigma_y^2(T_0/2).
\end{equation}
In the case of flicker noise, the corresponding coherence time $T_0$ is longer 
by a factor of 4 
compared to the case where the average frequency is predicted from past values 
only (Eq.\ (\ref{eq:varphi_adev})).

\subsection*{Coherence time in radio astronomy}

In radio astronomy the coherence time $T_\mathrm{co}$ of an oscillator with 
frequency $\nu_0$ is commonly estimated from the standard Allan deviation 
$\sigma_y$ \cite{kle72,rog81,tho07b}:

\begin{equation}\label{eq:tau_coh}
2\pi\nu_0\, T_\mathrm{co}\,\sigma_y(T_\mathrm{co})=1\:.
\end{equation}

According to Eq.\ (\ref{eq:varphi_adev}) this definition corresponds to 
$\Delta\phi_\mathrm{rms} = \sqrt{2}$ \footnote{The radio astronomical 
definition of coherence time might originate from a different, not commonly 
used definition of the Allan deviation without a factor of $1/2$ leading to 
Eq.\ (\ref{eq:tau_coh}) \cite{kle72,rog81}}.

From the measured flicker floor of  $\mathrm{mod}~\sigma_y = 4 \times 10^{-17}$ 
(corresponding to standard Allan deviation $\sigma_y = 5 \times 10^{-17}$)  
we thus find a coherence time of about 16~s which is larger by a factor of 
$\sqrt{2}$ compared to the value defined by Eq.\ (\ref{eq:sigma_y}).

\bibliographystyle{apsrev4-1}
%

\end{document}